\documentclass[superscriptaddress,preprintnumbers,amsmath,amssymb,twocolumn,floats,aps,prl,reprint]{revtex4-1}
\usepackage{txfonts}
\usepackage{amssymb}
\usepackage{graphicx}
\usepackage{sidecap}
\usepackage{CJK}
\usepackage{txfonts}
\usepackage{url}
\usepackage[]{sidecap}

\begin{document}

\title{Roles of the Narrow Electronic Band near the Fermi Level in 1\textit{T}-TaS$_2$-Related Layered Materials}

\author{Chenhaoping Wen}
\affiliation{School of Physical Science and Technology, ShanghaiTech University, Shanghai 201210, China}

\author{Jingjing Gao}
\affiliation{Key Laboratory of Materials Physics, Institute of Solid State Physics, Chinese Academy of Sciences, Hefei 230031, China}
\affiliation{University of Science and Technology of China, Hefei 230026, China}

\author{Yuan Xie}
\affiliation{School of Physical Science and Technology, ShanghaiTech University, Shanghai 201210, China}

\author{Qing Zhang}
\author{Pengfei Kong}
\affiliation{School of Physical Science and Technology, ShanghaiTech University, Shanghai 201210, China}

\author{Jinghui Wang}

\affiliation{School of Physical Science and Technology, ShanghaiTech University, Shanghai 201210, China}
\affiliation{ShanghaiTech Laboratory for Topological Physics, ShanghaiTech University, Shanghai 201210, China}

\author{Yilan Jiang}
\affiliation{School of Physical Science and Technology, ShanghaiTech University, Shanghai 201210, China}

\author{Xuan Luo}
\affiliation{Key Laboratory of Materials Physics, Institute of Solid State Physics, Chinese Academy of Sciences, Hefei 230031, China}

\author{Jun Li}
\affiliation{School of Physical Science and Technology, ShanghaiTech University, Shanghai 201210, China}
\affiliation{ShanghaiTech Laboratory for Topological Physics, ShanghaiTech University, Shanghai 201210, China}

\author{Wenjian Lu}
\affiliation{Key Laboratory of Materials Physics, Institute of Solid State Physics, Chinese Academy of Sciences, Hefei 230031, China}

\author{Yu-Ping Sun}
\email{ypsun@issp.ac.cn}
\affiliation{Key Laboratory of Materials Physics, Institute of Solid State Physics, Chinese Academy of Sciences, Hefei 230031, China}
\affiliation{High Magnetic Field Laboratory, Chinese Academy of Sciences, Hefei 230031, China}
\affiliation{Collaborative Innovation Centre of Advanced Microstructures, Nanjing University, Nanjing 210093, China}

\author{Shichao Yan}
\email{yanshch@shanghaitech.edu.cn}
\affiliation{School of Physical Science and Technology, ShanghaiTech University, Shanghai 201210, China}
\affiliation{ShanghaiTech Laboratory for Topological Physics, ShanghaiTech University, Shanghai 201210, China}

\date{\today}

\begin{abstract}
Here we use low-temperature scanning tunneling microscopy and spectroscopy to reveal the roles of the narrow electronic band in two 1$T$-TaS$_2$ related materials (bulk 1$T$-TaS$_2$ and 4$H_{\rm b}$-TaS$_2$). 4$H_{\rm b}$-TaS$_2$ is a superconducting compound with alternating 1$T$-TaS$_2$ and 1$H$-TaS$_2$ layers, where the 1$H$-TaS$_2$ layer has weak charge density wave (CDW) pattern and reduces the CDW coupling between the adjacent 1$T$-TaS$_2$ layers. In the 1$T$-TaS$_2$ layer of 4$H_{\rm b}$-TaS$_2$, we observe a narrow electronic band located near Fermi level, and its spatial distribution is consistent with the tight-binding calculations for two-dimensional 1$T$-TaS$_2$ layers. The weak electronic hybridization between the 1$T$-TaS$_2$ and 1$H$-TaS$_2$ layers in 4$H_{\rm b}$-TaS$_2$ shifts the narrow electronic band to be slightly above the Fermi level, which suppresses the electronic correlation induced band splitting. In contrast, in bulk 1$T$-TaS$_2$, there is an interlayer CDW coupling induced insulating gap. In comparison with the spatial distributions of the electronic states in bulk 1$T$-TaS$_2$ and 4$H_{\rm b}$-TaS$_2$, the insulating gap in bulk 1$T$-TaS$_2$ results from the formation of a bonding band and an antibonding band due to the overlap of the narrow electronic bands in the dimerized 1$T$-TaS$_2$ layers.

\end{abstract}

\maketitle
Atomic layers with narrow electronic band near Fermi level offer unique platforms for the emergence of unconventional superconducting or magnetic phases~\cite{Cao2018, Sharpe2019, Lu2019}. In the narrow electronic band, the kinetic energy of the electrons is strongly reduced. When the Fermi level of the electronic system lies within the narrow band, the Coulomb interactions between electrons can significantly exceed their kinetic energy and drives the system into correlated electronic phases. When the atomic layers with narrow electronic band stack together, the interlayer coupling between the adjacent stacking layers can also strongly modify their narrow electronic bands~\cite{Yankowitz2019, Ritschel_PRB_2018, YDWang_NC_2020}. The interplay between in-plane electron-electron interactions and interlayer coupling makes the electronic properties of layered materials with narrow electronic band near Fermi level extremely complicated.

\begin{figure*}[t]
\centering
\includegraphics[width = 130mm]{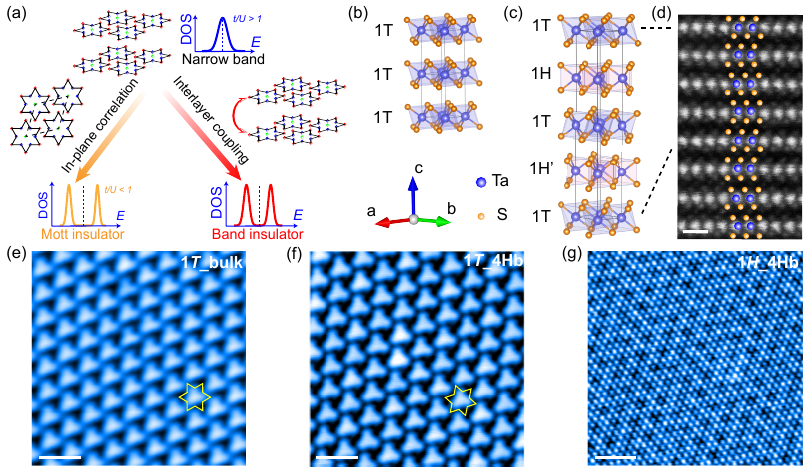}
\caption{(a) Schematic showing the influences of in-plane electron-electron interactions and interlayer coupling in the electronic states of 1$T$-TaS$_2$ layers with narrow electronic band. (b) Schematic for the structure of bulk 1$T$-TaS$_2$. (c) Schematic for the structure of 4$H_{\rm b}$-TaS$_2$. (d) High-resolution transmission electron microscopy image of 4$H_{\rm b}$-TaS$_2$ which shows the alternate stacking of the 1$T$-TaS$_2$ and 1$H$-TaS$_2$ layers. Scale bar: 0.5~nm. (e) Constant-current STM topography taken on bulk 1$T$-TaS$_2$ ($V_s$ = $-$500~mV, $I$ = 50~pA). (f) Constant-current STM topography taken on the 1$T$\textunderscore4$H_{\rm b}$ ($V_s$ = $-$500~mV, $I$ = 20~pA). (g) Constant-current STM topography taken on the 1$H$\textunderscore4$H_{\rm b}$ ($V_s$ = $-$500~mV, $I$ = 50~pA). Scale bar for (e), (f), and (g): 2~nm. }
\label{Sample}
\end{figure*}

One prominent example about the important and complicated roles of the narrow electronic band near Fermi level is the origin of the low-temperature insulating phase in bulk 1$T$-TaS$_2$ which has been debated over the past 40 years~\cite{Ritschel_PRB_2018, YDWang_NC_2020, Kim_PRL_1994, Fazekas_PMB_1979, Butler_NC_2020, Ritschel_NP_2015, Lee_PRL_2019}. Below 350~K, the in-plane lattice distortion leads to the formation of Star-of-David (SD) cluster which consists of 13 Ta atoms~\cite{Wilson_AIP_1975}. There is a single Ta-5d electron in each Ta site and each SD cluster has an odd number of electrons. Below 180~K, the SD clusters become long-range ordered and the bulk 1$T$-TaS$_2$ enters the commensurate $\sqrt{13}\times\sqrt{13}$ charge density wave (CDW) state with an insulating ground state~\cite{Rossnagel_JPCM_2011}. Without considering interlayer CDW coupling, the tight-binding simulations for two-dimensional 1$T$-TaS$_2$ indicate there is a very narrow electronic band near Fermi level in the commensurate CDW state of 1$T$-TaS$_2$ and it may be susceptible to Mott-Hubbard transition~\cite{Rossnagel_PRB_2006, Smith_JOPC_1985}. The measured low-temperature insulating gap in bulk 1$T$-TaS$_2$ is first interpreted as a Mott-insulating gap due to the strong electron-electron interactions [Fig.~\ref{Sample}(a)]~\cite{Kim_PRL_1994, Cho2016, Cho_PRB_2015}. However, the recent scanning tunneling microscopy (STM) and angle-resolved photoemission spectroscopy (ARPES) studies demonstrate that the interlayer CDW dimerization doubles the unit cell to contain even number of Ta-5d-orbital electrons, and the insulating gap in bulk 1$T$-TaS$_2$ is a band insulating gap [Fig.~\ref{Sample}(a)]~\cite{Ritschel_PRB_2018, YDWang_NC_2020, Butler_NC_2020, Ritschel_NP_2015, Lee_PRL_2019, Ma2016}. In this scenario, studying the electronic properties of the 1$T$-TaS$_2$ layers with strongly reduced interlayer CDW coupling would be very helpful to further reveal the roles of the narrow electronic band in the 1$T$-TaS$_2$ layer.

4$H_{\rm b}$-TaS$_2$ is another 1$T$-TaS$_2$ related layered material in which the unit cell consists of alternating layers of 1$T$-TaS$_2$ and 1$H$-TaS$_2$ (half of 2$H$-TaS$_2$) [Fig.~\ref{Sample}(c)]. The alternate stacking of the 1$T$-TaS$_2$ and 1$H$-TaS$_2$ layers can be clearly seen in the high-resolution transmission electron microscopy image [Fig.~\ref{Sample}(d)]. In contrast to the low-temperature insulating phase in bulk 1$T$-TaS$_2$, 4$H_{\rm b}$-TaS$_2$ is a superconductor with transition temperature $T$$_c$ $\sim$ 3 K (see Supplemental Materials Fig.S1,~\cite{Supplemental}). Although the $\sqrt{13}\times\sqrt{13}$ CDW order exists in the 1$T$-TaS$_2$ layer of 4$H_{\rm b}$-TaS$_2$ (1$T$\textunderscore4$H_{\rm b}$), the CDW order in the 1$H$-TaS$_2$ layer of 4$H_{\rm b}$-TaS$_2$ (1$H$\textunderscore4$H_{\rm b}$) is very weak~\cite{Ekvall_PRB_1997, Han_PRB_1994, Kim_PRB_1995}. The 1$H$\textunderscore4$H_{\rm b}$ can greatly reduce the interlayer CDW coupling between adjacent 1$T$\textunderscore4$H_{\rm b}$, and the previous ARPES measurements indicate 1$T$\textunderscore4$H_{\rm b}$ and 1$H$\textunderscore4$H_{\rm b}$ still retain their original electronic dispersion~\cite{Ribak_ScienceAdvances_2020}. This makes 4$H_{\rm b}$-TaS$_2$ an interesting platform to investigate the electronic properties of the 1$T$-TaS$_2$ layers with strongly reduced interlayer CDW coupling.

\begin{figure}[htb]
    \includegraphics[width = 78mm]{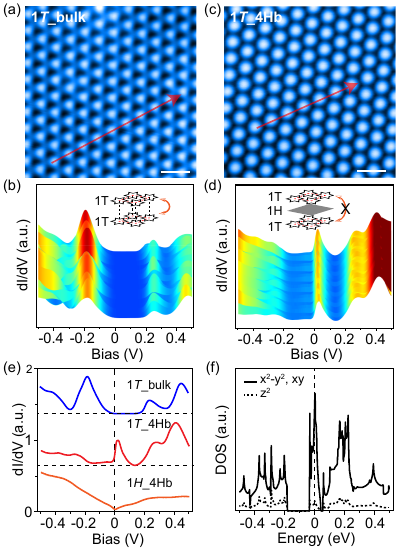}
    \caption{(a) and (c) Constant-current STM topographies taken on the bulk 1$T$-TaS$_2$ (a) and on the 1$T$\textunderscore4$H_{\rm b}$ (c) (a: $V_s$ = 500~mV, $I$ = 100~pA; c: $V_s$ = 500~mV, $I$ = 20~pA). Scale bar for (a) and (c): 2~nm. (b) and (d) Spatially distributed d$I$/d$V$ spectra along the arrows shown in (a) and (c), respectively. The insets in (b) and (d) are the sketches showing the interlayer CDW coupling in bulk 1$T$-TaS$_2$ and the reduced interlayer CDW coupling in 4$H_{\rm b}$-TaS$_2$. (e) Typical d$I$/d$V$ spectra taken on the bulk 1$T$-TaS$_2$ (blue), the 1$T$\textunderscore4$H_{\rm b}$ (red) and the 1$H$\textunderscore4$H_{\rm b}$ (orange). Setup condition: 500~mV, 1~nA. The dashed horizontal lines indicate d$I$/d$V$ = 0 for each spectrum. (f) The previously calculated orbital-projected density of states for two-dimensional 1$T$-TaS$_2$~\cite{Rossnagel_PRB_2006}. }
\label{Spectra}
\end{figure}

In this letter, we use low-temperature STM to probe and reveal the roles of the narrow electronic band in the low-temperature electronic states of 4$H_{\rm b}$-TaS$_2$ and bulk 1$T$-TaS$_2$. We find that the narrow electronic band near Fermi level exists in the 1$T$\textunderscore4$H_{\rm b}$ and it is localized near the center of the SD cluster. We show that the correlation effect in the 1$T$\textunderscore4$H_{\rm b}$ depends on the filling factor of the narrow electronic band. We also demonstrate the relationship between the narrow electronic band in 1$T$-TaS$_2$ layer and the low-temperature insulating gap in bulk 1$T$-TaS$_2$.

Bulk 1$T$-TaS$_2$ and 4$H_{\rm b}$-TaS$_2$ single crystals were synthesized by chemical vapor transport method with iodine as a transport agent and the detailed growth conditions were described elsewhere~\cite{Gao_PRB_2020, Shen_NanoLetters_2020}. STM experiments were performed with a home-built low-temperature STM. Bulk 1$T$-TaS$_2$ and 4$H_{\rm b}$-TaS$_2$ samples were cleaved at 77~K and then transferred into the low-temperature STM head for measurements at 4.3~K which is slightly above the superconducting transition temperature of 4$H_{\rm b}$-TaS$_2$. Chemically etched tungsten tips were flashed by electron-beam bombardment for several minutes before use. Scanning tunneling spectroscopy (STS) measurements were done by using standard lock-in technique with 3~mV modulation at the frequency of 914~Hz.

Figure~\ref{Sample}(e) and (f) are the negative-bias-voltage STM topographies taken on the bulk 1$T$-TaS$_2$ and the 1$T$\textunderscore4$H_{\rm b}$, respectively. The $\sqrt{13}\times\sqrt{13}$ CDW pattern can be clearly seen on both these surfaces. In the negative-bias-voltage STM topography taken on the 1$H$\textunderscore4$H_{\rm b}$, the intrinsic $3\times3$ CDW pattern can be barely seen [Fig.~\ref{Sample}(g)]. Due to the CDW order in the underneath 1$T$\textunderscore4$H_{\rm b}$, there is a modulation pattern with the $\sqrt{13}\times\sqrt{13}$ periodicity in the positive-bias-voltage STM topography taken on the 1$H$\textunderscore4$H_{\rm b}$ (Supplemental Materials Figs.S2 and S3,~\cite{Supplemental}). Our STM topographies indicate that the CDW pattern in the 1$H$\textunderscore4$H_{\rm b}$ is weak, which makes it behaves as a buffer layer between the adjacent 1$T$\textunderscore4$H_{\rm b}$.

We next perform STS to study and compare the local electronic structures in bulk 1$T$-TaS$_2$ and 1$T$\textunderscore4$H_{\rm b}$. Figure~\ref{Spectra}(a) is the STM topography taken on the most typical cleave plane of bulk 1$T$-TaS$_2$. Figure~\ref{Spectra}(b) shows the line cut differential conductance (d$I$/d$V$) spectra along the red arrow shown in Fig.~\ref{Spectra}(a), where the $\sim$150~mV insulating gap induced by the interlayer dimerization effect can be clearly seen~\cite{Butler_NC_2020}. In comparison, Figure~\ref{Spectra}(d) shows the line cut d$I$/d$V$ spectra taken on the 1$T$\textunderscore4$H_{\rm b}$ along the red arrow shown in Fig.~\ref{Spectra}(c). As shown in Fig.~\ref{Spectra}(d), there is a sharp peak feature located just above the Fermi level, and the peak width at half height is $\sim$50~mV. Another clear feature in Fig.~\ref{Spectra}d is a gap-like feature between $-$200mV and the Fermi level. Figure~\ref{Spectra}(e) shows the typical d$I$/d$V$ spectra taken on the 1$H$\textunderscore4$H_{\rm b}$, 1$T$\textunderscore4$H_{\rm b}$ and the bulk 1$T$-TaS$_2$. The overall feature of the d$I$/d$V$ spectrum on 1$H$\textunderscore4$H_{\rm b}$ is $V$-shaped without clear electronic peaks and gaps (Supplemental Materials Figs.S4 and S5, ~\cite{Supplemental}).

\begin{figure*}
    \includegraphics[width = 115mm]{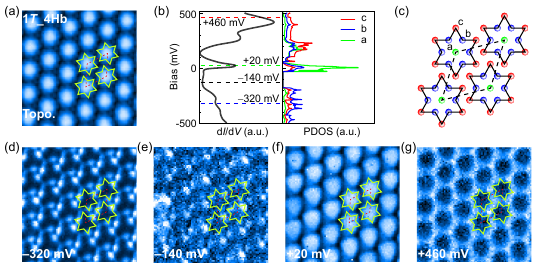}
    \caption{(a) Constant-current STM topography taken on 1$T$\textunderscore4$H_{\rm b}$ ($V_s$ = 500~mV, $I$ = 1~nA). Size: 6~nm $ \times$ 6~nm. (b) The typical d$I$/d$V$ spectrum taken on 1$T$\textunderscore4$H_{\rm b}$ (left panel). The right panel shows the previously calculated atom-projected density of states on the `a', `b' and `c' Ta atoms shown in (c) ~\cite{Rossnagel_PRB_2006}. (c) Unit-cell of the $\sqrt{13}\times\sqrt{13}$ reconstruction in the 1$T$-TaS$_2$ layer. The green, blue and red dots denote the inequivalent Ta atoms in the SD clusters. (d)-(g) d$I$/d$V$ maps taken on the same region as shown in (a) with $-$320~mV (d), $-$140~mV (e), $+$20~mV (f) and $+$460~mV (g) bias voltages. }
\label{QPI_4Hb}
\end{figure*}

Because of the 1$H$\textunderscore4$H_{\rm b}$, the interlayer CDW coupling between the adjacent 1$T$\textunderscore4$H_{\rm b}$ is reduced. To understand the features in the d$I$/d$V$ spectrum taken on the 1$T$\textunderscore4$H_{\rm b}$, we compare it with the previous empirical tight-binding simulations for two-dimensional 1$T$-TaS$_2$ without considering the interlayer coupling [Fig.~\ref{Spectra}(f)]~\cite{Rossnagel_PRB_2006}. Surprisingly, the overall feature in the d$I$/d$V$ spectrum of the 1$T$\textunderscore4$H_{\rm b}$ matches with the calculated density of states shown in Fig.~\ref{Spectra}(f), including the narrow band near Fermi level and the gap between $-$200~mV and the Fermi level. The difference is that the whole spectrum measured on the 1$T$\textunderscore4$H_{\rm b}$ is shifted up by $\sim$20~mV [Fig.~\ref{Spectra}(e)]. This is likely due to the weak electronic hybridization between the 1$T$\textunderscore4$H_{\rm b}$ and the 1$H$\textunderscore4$H_{\rm b}$, which induces effective doping effect to the 1$T$\textunderscore4$H_{\rm b}$~\cite{Shao2019}. This is also consistent with the previous photoemission measurements on 4$H_{\rm b}$-TaS$_2$ (Supplemental Materials Fig. S8, ~\cite{Supplemental})~\cite{Hughes_PRL_1995,Ribak_ScienceAdvances_2020}.

\begin{figure*}
    \centering
    \includegraphics[width = 115mm]{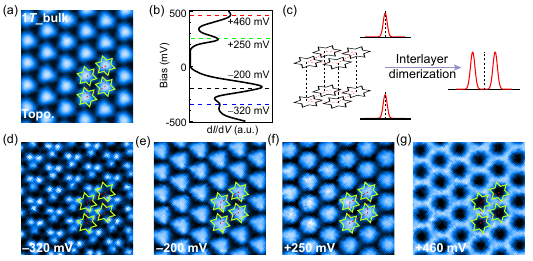}
    \caption{(a) Constant-current STM topography taken on bulk 1$T$-TaS$_2$ ($V_s$ = 500~mV, $I$ = 1~nA). Size: 6~nm $\times$ 6~nm. (b) The typical d$I$/d$V$ spectrum taken on bulk 1$T$-TaS$_2$. (c) Schematic showing the interlayer CDW dimerization induced band insulating gap in bulk 1$T$-TaS$_2$. (d)-(g) d$I$/d$V$ maps taken on the same region as shown in (a) with $-$320~mV (d), $-$200~mV (e), $+$250~mV (f) and $+$460~mV (g) bias voltages.}
\label{QPI_1T}
\end{figure*}

To further prove that the measured narrow electronic band near the Fermi level in the 1$T$\textunderscore4$H_{\rm b}$ has the same characters as the calculated narrow electronic band in the previous tight-binding simulations [Fig.~\ref{Spectra}(f)]~\cite{Rossnagel_PRB_2006}, we perform spatially resolved STS to map its spatial distribution and compare with the atom-projected tight-binding simulations. As shown in Fig.~\ref{QPI_4Hb}(c), there are three kinds of inequivalent Ta atoms in each SD cluster, which are labelled as `a', `b', and `c' atoms. According to the tight-binding simulations [right panel in Fig.~\ref{QPI_4Hb}(b)]~\cite{Rossnagel_PRB_2006}, in the energy range of $-$500~mV and $+$500~mV, only the calculated narrow electronic band near the Fermi level has higher intensity in the central `a' atom of the SD cluster. Figures~\ref{QPI_4Hb}(d)-(g) are the d$I$/d$V$ maps taken on the same region shown in Fig.~\ref{QPI_4Hb}(a) with $-$320~mV, $-$140~mV, $+$20~mV, and $+$460~mV bias voltages, respectively. The CDW unit cell as well as the SD clusters are also outlined in each d$I$/d$V$ map. As shown in Figs.~\ref{QPI_4Hb}(d)-(g), the narrow electronic band located near Fermi level prefers to locate in the center of the SD cluster [Fig.~\ref{QPI_4Hb}(f)] and all the other electronic bands have higher intensity at the rim of the SD cluster, which agree very well with the atom-projected density of states in the tight-binding simulations [right panel in Fig.~\ref{QPI_4Hb}(b)]~\cite{Rossnagel_PRB_2006}.

After characterizing the narrow electronic band in the 1$T$\textunderscore4$H_{\rm b}$, we try to explore the relationship between this narrow electronic band and the insulating gap in the bulk 1$T$-TaS$_2$. We also use spatially resolved STS to measure the spatial distributions of the electronic states in bulk 1$T$-TaS$_2$. As shown in Figs.~\ref{QPI_1T}(d) and (g), the spatial distributions of the electronic states at $-$320~mV and $+$460~mV are consistent with the corresponding electronic states in the 1$T$\textunderscore4$H_{\rm b}$ [Figs.~\ref{QPI_4Hb}(d) and (g)] which prefer to locate at the rim the SD cluster. Interestingly, the spatial distributions of the electronic peaks below and above the insulating gap [Figs.~\ref{QPI_1T}(e) and (f)] agree with the spatial distribution of the narrow electronic band in the 1$T$\textunderscore4$H_{\rm b}$ [Fig.~\ref{QPI_4Hb}(f)], which prefer to reside on the center of the SD cluster~\cite{PRX2017}. The recent STM work has demonstrated that the $\sim$150~meV insulating gap in bulk 1$T$-TaS$_2$ [Fig.~\ref{QPI_1T}(b)] appears when the CDW patterns are vertical aligned in the dimerized 1$T$-TaS$_2$ layers [Fig.~\ref{QPI_1T}(c)]~\cite{Butler_NC_2020}. Our data further indicates that in that stacking configuration, the narrow electronic bands in the dimerized 1$T$-TaS$_2$ layers have strong spatial overlap and form the bonding and antibonding bands [Fig.~\ref{QPI_1T}(c)].

We next discuss why the narrow electronic band in the 1$T$\textunderscore4$H_{\rm b}$ is not splitted by the in-plane electron-electron interactions. This is likely because of the weak electronic hybridization between the 1$T$\textunderscore4$H_{\rm b}$ and 1$H$\textunderscore4$H_{\rm b}$ which shifts the narrow electronic band to be slightly above the Fermi level. This makes the narrow electronic band in the 1$T$\textunderscore4$H_{\rm b}$ almost unfilled, and suppresses the correlation induced electronic gap. The filling-factor-dependent correlation gap has been reported in twisted-bilayer graphene, where the correlation induced band splitting strongly depends on the filling factor of the flat electronic band~\cite{Jiang_Nature_2019}. However, the electronic hybridization is usually sensitive to the interlayer distances between the two layers. In the 4$H_{\rm b}$-TaS$_2$ sample, we find that in a few regions the narrow electronic band is located slightly closer to the Fermi level, and a $\sim$30~meV electronic gap can be detected (see Supplemental Materials Fig.9,~\cite{Supplemental}). This $\sim$30~mV gap may be due to the electronic-correlation-induced gap in the 1$T$\textunderscore4$H_{\rm b}$. This also agrees with the size of the possible correlation gap ($\sim$50~meV) in the undimerized 1$T$-TaS$_2$ layer of bulk 1$T$-TaS$_2$~\cite{Butler_NC_2020}. For the further experiments, it would be interesting to investigate correlation effect in the 1$T$\textunderscore4$H_{\rm b}$ by moving the narrow electronic band closer to Fermi level with electron doping.

In summary, our data not only shows the existence of the narrow electronic band near Fermi level in 1$T$-TaS$_2$ layer, but also demonstrates it plays a crucial role in the low-temperature electronic properties of 1$T$-TaS$_2$ related materials. When the narrow electronic band in the 1$T$\textunderscore4$H_{\rm b}$ is located slightly above the Fermi level, the electronic correlation induced narrow band splitting is suppressed. The insulating gap in bulk 1$T$-TaS$_2$ originates from the interlayer CDW dimerization induced bonding and antibonding bands of the narrow electronic band. Our work also paves the way for further understanding the novel electronic phases in 1$T$-TaS$_2$ related materials, such as the possible chiral superconductivity in 4$H_{\rm b}$-TaS$_2$~\cite{Ribak_ScienceAdvances_2020} and the electronic states in single layer 1$T$-TaS$_2$~\cite{PRB_2014,Lin_Nano_Research_2019}.

S.Y. acknowledges the financial support from Science and Technology Commission of Shanghai Municipality (STCSM) (Grant No. 18QA1403100), National Science Foundation of China (Grant No. 11874042) and the start-up funding from ShanghaiTech University. J.L. acknowledges the financial support from National Science Foundation of China (Grant No. 61771234). J.J.G., X.L., W.J.L and Y.P.S. thank the support of National Key Research and Development Program under Contract No. 2016YFA0300404, the National Nature Science Foundation of China under Contracts No. 11674326 and No. 11774351, and the Joint Funds of the National Natural Science Foundation of China and the Chinese Academy of Sciences' Large-Scale Scientific Facility under Contracts No. U1832141, No. U1932217 and U2032215. TEM experiments were supported by ShanghaiTech University and the Center for High-resolution Electron Microscopy (C$\hbar$EM) through EM-19430216.

\end{document}